\documentclass[aip,amsmath,amssymb,reprint]{revtex4-1}
\usepackage{graphicx}
\usepackage{amsmath}
\usepackage{bm}
\usepackage{dcolumn}
\usepackage[utf8]{inputenc}
\usepackage[T1]{fontenc}
\usepackage{mathptmx}
\usepackage{color}
\definecolor{green}{rgb}{0,0.7,0}

\begin{document}

\title{On Richtmyer-Meshkov unstable dynamics of three-dimensional \\ interfacial coherent structures with time-dependent acceleration}

\author{D.L.~Hill}
\email{des.hill@uwa.edu.au}
\affiliation{University of Western Australia, Perth,WA, 6009, Australia}

\author{S.I.~Abarzhi}
\email{snezhana.abarzhi@gmail.com}
\affiliation{University of Western Australia, Perth,WA, 6009, Australia}

\date{\today}

\begin{abstract}
Richtmyer-Meshkov instability (RMI) plays an important role in many areas of science and engineering, from supernovae
and fusion to scramjets and nano-fabrication. Classical Richtmyer-Meshkov instability is induced by a steady shock and impulsive
acceleration, whereas in realistic environments the acceleration is usually variable. We focus on RMI induced
by acceleration with power-law time-dependence and apply group theory to solve the long-standing problem. For early-time
dynamics, we find the dependence of the growth-rate on the initial conditions and show that it is independent of the acceleration
parameters. For late-time dynamics, we find a continuous family of regular asymptotic solutions, including their curvature,
velocity, Fourier amplitudes, and interfacial shear, and we study their stability.  For each solution, the interface dynamics
is directly linked to the interfacial shear, the non-equilibrium velocity field has intense fluid motion near the interface and effectively
no motion in the bulk. The quasi-invariance of the fastest stable solution suggests that nonlinear coherent dynamics in RMI is
characterized by two macroscopic length-scales - the wavelength and the amplitude, in agreement with observations. The properties
of a number of special solutions are outlined, these being respectively, the Atwood,  Taylor, convergence, minimum-shear, and critical
bubbles, among others. We also elaborate new theory benchmarks for future experiments and simulations.
\end{abstract}

\maketitle

\section{Introduction}

Rayleigh-Taylor instability (RTI) develops at the fluid interface when fluids of different densities are accelerated against their
density gradients; Richtmyer-Meshkov instability (RMI) develops when the acceleration is induced by a shock and is impulsive
\cite{richtmyer,meshkov,rayleigh,davies}. Intense interfacial Rayleigh-Taylor (RT) / Richtmyer-Meshkov (RM) mixing of the fluids
ensues with time \cite{abarzhireview,meshkovbook}. RTI/RMI and RT/RM mixing play an important role in a broad range of
processes in nature and technology, including stellar evolution and plasma fusion, and in the fossil fuel industry
\cite{turbulentmixing,arnett,zeldovichraizer,haan,peters,rana,buehler}. In this work we study the long-standing problem of RMI
with variable acceleration \cite{supernovae}. We employ group theory to solve the boundary value problem for the early-time and
late-time RMI \cite{dynamicsreview}, directly link RM dynamics to the interfacial shear, identify its invariance properties,
and reveal the interfacial and multi-scale character of RM dynamics. Our theory finds similarities and differences between
RM and RT dynamics with variable accelerations \cite{supernovae}, agrees with existing observations, and elaborates
new diagnostic benchmarks for experiment and simulation.

RMI with variable acceleration commonly occur in fluids, plasmas, and materials
\cite{turbulentmixing,arnett,zeldovichraizer,haan,peters,rana,buehler,supernovae}: RMI leads to the appearance of light-years-long
structures in clouds of molecular hydrogen, influences the formation of hot spots in inertial confinement fusion, controls combustion
processes in scramjets, and drives material transformation under impact in nano-fabrication. In these vastly different physical
conditions, RM flows have similar qualitative features of their evolution. The post-shock RM dynamics is a superposition of two
motions. These are the background motion of the fluid bulk and the growth of the interface perturbations
\cite{richtmyer,meshkov,meshkovbook,stanic,dell2015,dell2017,psa}. In the background motion, both fluids and their interface
move as a whole in the direction of the transmitted shock. This motion occurs even for an ideally planar interface and is supersonic
for strong shocks. The growth of the interface perturbations is due to impulsive acceleration by the shock; it develops only when
the flow fields are perturbed; its growth rate is subsonic and the associated motion is incompressible
\cite{richtmyer,meshkov,meshkovbook,stanic,dell2015,dell2017,psa,meshkov2013,robey,lugomer,swisher}. The growth rate is
constant initially and decays with time later. The RM unstable interface is transformed to a composition of small-scale shear-driven
vortical structures and a large-scale coherent structure of bubbles and spikes, where a bubble (spike) is a portion of the
light (heavy) fluid penetrating the heavy (light) fluid. Small-scale non-uniform structures appear also in the bulk, including hot and
cold spots, high and low pressure regions, cumulative jets, checker-board velocity patterns \cite{abarzhireview,meshkovbook,stanic,dell2015,dell2017,psa,meshkov2013}. Over time, self-similar RM mixing develops, and
energy supplied initially by the shock gradually dissipates
\cite{abarzhireview,meshkovbook,stanic,dell2015,dell2017,psa,meshkov2013,robey,lugomer,swisher,dynamicsreview,supernovae}.

RMI/RTI and RM/RT mixing are a challenge to study in theory, experiments and simulations
\cite{supernovae,dynamicsreview,stanic,dell2015,dell2017,psa,meshkov2013,robey,lugomer,swisher,orlov,anisimov,kull,
abarzhisteady,nishihara,bhowmick2016,gauthier,kadau,glimm,youngs,thornber}. As regards the general theory, we have to
develop new approaches for non-equilibrium multi-scale RM/RT dynamics, capture symmetries of these RM/RT dynamics
and identify properties of their asymptotic solutions
\cite{abarzhireview,dynamicsreview,anisimov,kull,abarzhisteady,nishihara,bhowmick2016,gauthier,supernovae}.
Experimental work requires one to meet tight requirements on the flow implementation, diagnostics and control
\cite{meshkov,meshkovbook,meshkov2013,robey,lugomer,swisher,orlov}. Simulations must employ highly accurate numerical
methods, requiring massive computations in order to capture shocks, track interfaces, and accurately model small-scale processes
\cite{stanic,dell2015,dell2017,psa,gauthier,kadau,glimm,youngs,thornber}. In addition, a substantial span of temporal and spatial
scales is required for bias-free interpretation of experimental and numerical data describing RM/RT evolution
\cite{supernovae,dynamicsreview,stanic,dell2015,dell2017,psa,meshkov2013,robey,lugomer,swisher,orlov,anisimov,kull,
abarzhisteady,nishihara,bhowmick2016,gauthier,kadau,glimm,youngs,thornber}. Significant success has recently been achieved
in our understanding of RMI and RTI, as well as RM and RT mixing \cite{abarzhireview,dynamicsreview,anisimov,supernovae}.
In particularly, the group theory approach has uncovered the multi-scale character of nonlinear RMI and RTI, and found an order
in RT mixing with constant acceleration, thus explaining observations \cite{abarzhireview,dynamicsreview,anisimov,supernovae}.
 
In realistic environments, RT and RM flows are usually driven by variable acceleration 
\cite{arnett,zeldovichraizer,haan,peters,rana,buehler,supernovae} and only limited information is currently available on
RM and RT dynamics under these conditions \cite{supernovae}. An important special case is that of acceleration with power-law
time-dependence, because power-law functions may lead to new scaling properties of the dynamics and can be used to adjust the acceleration's time-dependence in applications \cite{arnett,zeldovichraizer,haan,peters,rana,buehler,supernovae,landau,sedov}.
For such accelerations, the early-time and late-time scale-dependent dynamics can be of RM or RT type, depending on the
exponent in the acceleration power-law \cite{supernovae,hbia}. Specifically, the interfacial dynamics is driven by the acceleration
and is of RT-type for exponents larger than $(-2)$, and is driven by the initial growth-rate and is of RM-type otherwise
\cite{supernovae,hbia}. A self-similar mixing regime ensues over time \cite{abarzhireview,meshkovbook}.
 
In this work, we study the long-standing problem of RMI with variable acceleration for a three-dimensional spatially extended
periodic flow. We apply group theory with the spacial symmetry group of the square to solve the boundary value problem involving
boundary conditions at the interface and the outside boundaries, and the initial value problem
\cite{abarzhireview,supernovae,dynamicsreview,anisimov,hbia}. For early-time dynamics, we find the dependence of the RMI
growth-rate on the  initial  conditions and show that it is independent of the acceleration parameters. For late-time dynamics,
we directly link the interface dynamics  to the interfacial shear, find a continuous family of regular asymptotic solutions, and study
the stability of the solutions in this family. For each of the family of solutions, the perturbed velocity field has intense fluid motion
near the interface and effectively no motion in the bulk. We identify  parameters of a number of special solutions, these being
the Atwood bubble, which is flat and is the fastest, and respectively, the Taylor, convergence, minimum-shear, and critical bubbles.
In each case we give the curvature, velocity, Fourier amplitudes, and interfacial shear. The Atwood bubble has a quasi-invariance
property suggesting that nonlinear coherent RM dynamics is set by the interplay of two macroscopic length-scales - the wavelength
and the amplitude. Our theory agrees with existing observations, and elaborates new benchmarks for observations 
\cite{meshkovbook,arnett,zeldovichraizer,haan,peters,rana,buehler,supernovae,meshkov2013,robey,lugomer,swisher}.

\section{The method of solution}

\subsection{The governing equations}

The dynamics of ideal fluids is governed by conservation of mass, momentum and energy:
\[\frac{\partial \rho}{\partial t}+\frac{\partial}{\partial x_i}(\rho v_i)=0,\]
\[\frac{\partial}{\partial t}(\rho v_j)+\frac{\partial}{\partial x_i}(\rho v_iv_j)+\frac{\partial P}{\partial x_j}=0,\]
\begin{equation} \label{eq:laws}
\frac{\partial E}{\partial t}+\frac{\partial}{\partial x_i}((E+P)v_i)=0,
\end{equation}
where $j=1,2,3$, $(x_1,x_2,x_3)=(x,y,z)$ are the spatial coordinates, $t$ is time, ($\rho,{\bf v},P,E)$ are the
fields of density $\rho$ , velocity ${\bf v}$ , pressure $P$ and energy $E=\rho(e+{\bf v}^2/2)$, where
$e$ is the specific internal energy \cite{abarzhisteady}. The flow fields are understood as post-shock in the inertial frame of reference
moving with the velocity of the background motion, in the direction of the transmitted shock.

We consider immiscible, inviscid fluids of differing densities, separated by a sharp interface. It is required that  momentum must
be conserved at the interface and that there can be no mass flow across it. Hence the boundary conditions at the interface are
\begin{equation} \label{eq:ics}
\left[{\bf v}\cdot{\bf n}\right]=0, \quad [P]=0, \quad
\left[{\bf v}\cdot{\bm \tau}\right]={\rm arbitrary}, \quad \left[W\right]={\rm arbitrary},
\end{equation}
where $[\cdots]$ denotes the jump of functions across the interface; ${\bf n}$ and ${\bm \tau}$ are the normal and
tangential unit vectors of the interface with ${\bf n}={\bm \nabla}\theta/\vert{\bm \nabla}\theta\vert$ and
${\bf n}\cdot{\bm \tau}=0$; $W=e+P/\rho$ is the specific enthalpy; $\theta=\theta(x,y,z,t)$ is a local scalar
function, with $\theta=0$ at the interface and $\theta>0$ $(\theta<0)$ in the bulk of the heavy (light) fluid, indicated
hereafter by subscript $h(l)$.

The heavier fluid sits above the lighter fluid and the entire system is subject to a time-dependent acceleration field,
directed from the heavy to the light fluid (and transmitted in the direction of the shock). The acceleration is the power-law function of time,
${\bf g}=(0,0,-g)$ where $g=Gt^a$. Here $a$ is the acceleration exponent, and $G>0$ is the acceleration pre-factor \cite{nishihara,garabedian,inogamov}. Their dimensions are $[G]=ms^{-(a+2)}$ and $[a]=1$. This modifies the pressure field.

We assume that the outside boundaries do not influence the dynamics and there are no mass sources. Hence at the
outside boundaries of the domain, the boundary conditions are
\begin{equation} \label{eq:bcs}
\lim_{z\to\infty}{\bf v}_h={\bf 0}, \hspace{2cm} \lim_{z\to-\infty}{\bf v}_l={\bf 0}.
\end{equation}

\subsection{Large-scale coherent structures}

RM large-scale coherent structures are the arrays of bubbles and spikes periodic in the plane normal to the acceleration
direction. At large scales the flow can be assumed to be irrotational in the bulk.  We also assume that the
fluids are incompressible and hence that the velocities are expressible in terms of scalar potentials $\Phi_h(x,y,z,t)$
and $\Phi_l(x,y,z,t)$. Because the fluids are ideal these are harmonic, with $\nabla^2\Phi_h=0\ {\rm in}\ \theta>0$
and $\nabla^2\Phi_l=0$ in $\theta<0$.

For convenience we perform the calculations in the non-inertial frame of reference moving with velocity $v(t)$ in the
$z$-direction, where $v(t)=\partial z_0/\partial t$ and $z_0$ are the velocity and position (amplitude) in the inertial reference
frame at a regular point of the interface, such as the tip of a bubble or a spike. Then the interface function is
$\theta(x,y,z,t)=z-z^*(x,y,t)=0$, and the interface conditions are
\[\rho_h\left(\nabla\Phi_h\cdot{\bf n}+\frac{\dot\theta}{\vert\nabla\theta\vert}\right)=0
=\rho_l\left(\nabla\Phi_l\cdot{\bf n}+\frac{\dot\theta}{\vert\nabla\theta\vert}\right),\]
\[\rho_h\left(\frac{\partial \Phi _h}{\partial t}+\frac{\vert\nabla\Phi_h\vert^2}{2}
+\left(g(t)+\frac{dv}{dt}\right)z\right)\]
\[=\rho_l\left(\frac{\partial \Phi _l}{\partial t}+\frac{\vert\nabla\Phi_l\vert^2}{2}
+\left(g(t)+\frac{dv}{dt}\right)z\right),\]
\begin{equation} \label{eq:interfaceconds}
\nabla\Phi_h\cdot{\bm \tau}-\nabla\Phi_l\cdot{\bm \tau}={\rm arbitrary}
\end{equation}
In the non-inertial reference frame, the outside boundary condition (\ref{eq:bcs}) takes the form
\begin{equation} \label{eq:farfield}
\nabla\Phi _h\Big\vert_{z\to\infty}=(0,0,-v(t)), \hspace{1cm}
\nabla\Phi _l\Big\vert_{z\to\infty}=(0,0,-v(t)).
\end{equation}

\subsection{The dynamical system}

The length scale is $1/k$, where $k$ is a wavenumber, with $k=2\pi/\lambda$ and $\lambda$ being the wavelength.
There are two natural time scales in the problem \cite{hbia}. These are $\tau_g=(kG)^{-1/(a+2)}$ and $\tau_0=1/kv_0$,
where $v_0$ is some initial growth rate.
For $a<-2$, $\tau_0\ll\tau_G$ and the fastest process is set by the initial growth-rate; this is the initial growth-rate driven
Richtmyer-Meshkov type dynamics \cite{hbia}.  Hence, we set the time scale to be $\tau=\tau_0$ and we will consider the flow for
$t\gg t_0$ with $t_0\gg\tau$, that is, sufficiently later than the initial shock. The Atwood number
$A=(\rho_h-\rho_l)/(\rho_h+\rho_l)$ parametrises the ratio of densities of the fluids, and $0<A<1$.

The periodic nature of the large-scale coherent structure can be accommodated by appealing to the theory of space
groups \cite{anisimov,abarzhireview,dynamicsreview,supernovae}. As a specific example, we focus on three-dimensional flow
with square symmetry. The details of the procedure are given elsewhere
\cite{abarzhireview,dynamicsreview,supernovae,psa,anisimov,abarzhisteady,hbia,ang,hillba}. The symmetry group dictates
a specific Fourier series (an irreducible representation of the group) which can be used to solve the nonlinear boundary value
problem Eq. (\ref{eq:interfaceconds}), Eq. (\ref{eq:farfield}). We then make spatial expansions in the vicinity of the tip of a regular
point on the interface. This approach reduces the governing equations to a dynamical system of ordinary differential equations
in terms of interface variables and Fourier moments
\cite{anisimov,abarzhireview,dynamicsreview,inogamov,abarzhisteady,supernovae}.

The corresponding potentials are
\[\Phi_h(x,y,z,t)=\sum_{m,n=0}^\infty\Phi_{mn}(t)\left(\frac{\cos(mkx)\cos(nky)
e^{-\alpha_{mn}kz}}{\alpha_{mn}k}+z\right)\]
\[+f_h(t),\]
\[\Phi_l(x,y,z,t)=\sum_{m,n=0}^\infty\tilde\Phi_{mn}(t)\left(\frac{\cos(mkx)\cos(nky)
e^{\alpha_{mn}kz}}{\alpha_{mn}k}-z\right)\]
\begin{equation} \label{eq:phiforms}
+f_l(t),
\end{equation}
where $\alpha_{mn}=\sqrt{m^2+n^2}$, $m$ and $n$ are integers, $k=\frac{2\pi}{\lambda}$ is the wavenumber, 
$\Phi_{mn}$ and $\tilde\Phi_{mn}$ are the Fourier amplitudes for the heavy and light fluids respectively, with
$\Phi_{00}=\tilde\Phi_{00}=0$, and $f_h(t)$ and $f_l(t)$ are time-dependent functions. Symmetry requires that
$\Phi_{mn}=\Phi_{nm}$ and $\tilde\Phi_{mn}=\tilde\Phi_{nm}$. The sign of the $z$ term is determined by the
boundary condition Eq. (\ref{eq:farfield}).

For application of symmetry groups in Rayleigh-Taylor and Richtmyer-Meshkov instability, the reader is referred to
other works \cite{abarzhireview,dynamicsreview,supernovae,psa,anisimov,abarzhisteady,hbia,ang,hillba}.

In order to examine the local behavior of the interfacial dynamics in the vicinity of the bubble tip, we expand the
interface function in a power series in the vicinity of a regular point of the interface (e.g. the tip of the bubble or spike).
In the moving frame of reference, this is
\begin{equation} \label{eq:zetaform}
z^*(x,y,t)=\sum_{N=1}^\infty\sum_{i+j=1}^N\zeta_{ij}(t)x^{2i}y^{2j},
\end{equation}
where $\zeta_{ij}(t)=\zeta_{ji}(t)$ due to symmetry, $\zeta_1(t)=\zeta_{10}(t)$ is the the principal curvature at the
regular point, and $N=i+j$ is the order of the approximation. To lowest order (that is, $N=1$), the interface is
$z^*(x,y,t)=\zeta_1(t)(x^2+y^2)$.

The Fourier series and interface function are substituted into the governing equations and the resulting expressions
expanded as Taylor series. This yields a system of ordinary differential equations for $\Phi_m(t)$, $\tilde\Phi_m(t)$
and $\zeta_{ij}(t)$. We may express the potentials in terms of moments
\[M_{a,b,c}(t)=\sum_{mn}\Phi_{mn}(t)(mk)^a(nk)^b(\alpha_{mn}k)^c\]
and
\[\tilde M_{a,b,c}(t)=\sum_{mn}\tilde\Phi_{mn}(t)(mk)^a(nk)^b(\alpha_{mn}k)^c\]
We note that by symmetry, $M_{a,b,c}=M_{b,a,c}$ and $M_{a+2,b,c}+M_{a,b+2,c}=M_{a,b,c+2}$ and similarly
for $\tilde M$. At $N=1$, we abbreviate the series to second order in $x$ and $y$, and first order in $z$ since
$z^*(x,y,t)$ is quadratic in $x$ and $y$.

The boundary conditions at the interface and at the outside boundaries of the domain become
\begin{equation} \label{eq:kinematic}
\dot\zeta_1=4M_1\zeta_1+\frac{M_2}{2}, \hspace{1cm}
\dot\zeta_1=4\tilde M_1\zeta_1-\frac{\tilde M_2}{2},
\end{equation}
\begin{equation} \label{eq:momentum}
(1+A)\left(\frac{\dot M_1}{2}+\zeta_1\dot M_0-\frac{M_1^2}{2}\right)
=(1-A)\left(\frac{\dot {\tilde{M_1}}}{2}-\zeta_1\dot {\tilde M_0}-\frac{\tilde M_1^2}{2}\right),
\end{equation}
\begin{equation} \label{eq:momentshear}
M_1-\tilde M_1={\rm arbitrary}
\end{equation}
\begin{equation} \label{eq:farfieldeq}
M_0=-\tilde M_0=-v(t)
\end{equation}
where $M_0=M_{0,0,0}$, $M_1=M_{2,0,-1}$ and $M_2=M_{2,0,0}$. This representation in terms of moments
$M$ and $\tilde M$, and the interface variable $\zeta_1$, accommodates the nonlocal nature of the nonlinear
dynamics and enables us to investigate the interplay of harmonics and derive regular asymptotic solutions.

Our expressions can account for any number of harmonics in any order. Previous work with either $a=0$ or $G=0$
has demonstrated that the $N=1$ solutions properly capture the physical behaviour
\cite{anisimov,abarzhireview,dynamicsreview,garabedian,inogamov,abarzhisteady,supernovae,ang,abarzhireview}.
Hence we consider only the case when $N=1$.

\section{Results}

As previously mentioned, we are considering arrays of bubbles and spikes periodic in the plane normal to the acceleration
direction. Bubbles are intrusions of the lighter fluid into the heavier fluid and as such move upwards and are concave down.
Spikes are intrusions of the heavier fluid into the lighter fluid and as such move downwards and are concave up. We note that
the dynamics of bubbles is regular, whereas that of spikes is singular. Our early-time analysis applies in both cases.
Here, we focus our attention on the  later-time dynamics of bubbles. That of spikes will be discussed elsewhere.

\subsection{The early-time regime, $t-t_0\ll\tau$}

In the early-time regime, the system can be linearised and only first-order harmonics are needed, that is, the moments retain only
one Fourier amplitude.  The initial conditions at time $t_0$ are the initial curvature $\zeta_1(t_0)$ and velocity $v(t_0)$, and
$\vert v_0(t)\vert=v_0$.

For a broad class of initial conditions, integration of the governing equations is a challenge. The solution can be found
\cite{davies,landau,supernovae,hbia} when the amplitude of the initial perturbation is small $\tau k\vert v_0\vert\ll1$, and the
interface is nearly flat $\vert\zeta_1/k\vert\ll1$. The system reduces to
\begin{equation} \label{eq:earlytime}
\dot\zeta_1=\left(\frac{k^2}{4}\right)M_0, \hspace{1cm} \dot M_0=\frac{Ak}{2}M_0^2.
\end{equation}

When $t-t_0\ll\tau=1/kv_0$, only first order harmonics are retained in moments, that is, $M_0=2\Phi_{10}$,
$\tilde M_0=2\tilde\Phi_{10}$; $M_n=k^n\Phi_{10}$, $\tilde M_n=k^n\tilde\Phi_{10}$, $n=1,2$. For an almost flat interface the
solution is
\begin{equation} \label{eq:almostflat}
-\frac{\zeta}{k}=\frac{1}{2A}\ln\left(C_2\frac{t}{\tau}+C_1\right), \hspace{1cm}
v=\frac{4}{k}\frac{d}{dt}\left(-\frac{\zeta}{k}\right),
\end{equation}
where $C_1$ and $C_2$ are integration constants defined by the initial conditions $\zeta_0=\zeta(t_0)$
and $v_0=v(t_0)$ with $\zeta_0/k\ll1$ and $\tau k\vert v_0\vert\ll1$ \cite{davies,landau,supernovae}.
First-order analysis of the very-early-time ($t\sim\ t_0$) dynamics yields
\begin{equation} \label{eq:veryearlytime}
\zeta-\zeta_0\sim-\frac{k^2v_0}{4}(t-t_0), \hspace{1cm}
v-v_0\sim-\frac{Akv_0^2}{2}(t-t_0)
\end{equation}
which suggests that the positions of bubbles ($\zeta\le0,v\ge0$) and spikes ($\zeta\ge0,v\le0$) are defined by the initial
velocity field, with bubbles formed for $v(t_0)/v_0<0$ and spikes formed for $v(t_0)/v_0>0$. The
instability growth-rate is independent of the acceleration parameters , since the contributions of acceleration-induced terms
to early-time dynamics are negligible \cite{hillba}.

\subsection{The late-time regime, $t-t_0\gg\tau$}

In the later-time regime, spikes are singular (the singularity is finite-time), whereas bubbles are regular
\cite{anisimov,abarzhireview,dynamicsreview}. For $t\gg\tau$, higher order harmonics are retained in the expressions for the
moments, and regular asymptotic solutions are derived. We find asymptotic solutions for the relevant equations and determine
their stability. We assume asymptotic solutions of the form
\begin{equation} \label{eq:asymptforms}
\frac{\zeta_1}{k}\sim\left(\frac{t}{\tau}\right)^\alpha, \hspace{1cm}
\frac{M_n}{k^n},\ \frac{\tilde M_n}{k^n}\ \sim\ \frac{1}{k\tau}\left(\frac{t}{\tau}\right)^\gamma,
\end{equation}
where $\alpha$ and $\gamma$ are constants to be determined. Substitution into Eqs. (\ref{eq:kinematic}) and 
(\ref{eq:momentum}) lead, respectively, to the requirements that
\[\alpha-1=\alpha+\gamma=\gamma, \hspace{2cm} \gamma-1=2\gamma.\]
That is,
\[\alpha=0, \hspace{2cm} \gamma=-1.\]

We investigate the stability of these nonlinear asymptotic solutions by considering perturbations
\[\zeta_1(t)\to\zeta_1+\delta\zeta_1(t)\]
\[M_j(t)\to M_j(t)+\delta M_j(t), \quad \tilde M_j(t)\to \tilde M_j(t)+\delta \tilde M_j(t)\]
with
\begin{equation} \label{eq:symrequ}
\frac{\delta M_n}{M_n}\sim\frac{\tilde\delta M_n}{\tilde M_n}\sim\frac{\delta\zeta_1}{\zeta_1}
\sim\left(\frac{t}{\tau}\right)^\beta.
\end{equation}
Nonlinear asymptotic solutions are stable for ${\rm Re}[\beta]<0$ and are unstable otherwise.

Substituting the asymptotic forms Eq. (\ref{eq:symrequ}) into Eqs. (\ref{eq:kinematic}) and (\ref{eq:momentum}), employing a
dominant balance argument and solving the resulting set of equations, at $N=1$, we find a one-parameter family of solutions.
We choose the bubble curvature $\zeta_1$ to parameterise the family. The Fourier amplitudes are
\[\Phi_{10}=-\frac{2k+8\zeta_1}{3k+8\zeta_1}v, \quad \Phi_{20}=\frac{k+8\zeta_1}{6k+16\zeta_1}v,\]
\[\tilde\Phi_{10}=\frac{2k-8\zeta_1}{3k-8\zeta_1}v, \quad \tilde\Phi_{20}=\frac{-k+8\zeta_1}{6k-16\zeta_1}v,\]
\begin{equation} \label{eq:phisandv}
-v=2\Phi_{10}+2\Phi_{20}, \hspace{1cm} v=2\tilde\Phi_{10}+2\tilde\Phi_{20}.
\end{equation}
While these expressions are identical to those for the $a>-2$ case, and are repeated here for the convenience of the
reader, the velocity and shear solutions are very different from those obtained in the $a>-2$ case \cite{hillba}. Specifically,
the velocity is
\[\hat v=\frac{A(9-64\hat\zeta^2)(128A\hat\zeta^3-10A\hat\zeta+3)}{3(64A\hat\zeta^2+9A+48\hat\zeta)},\]
where
\begin{equation} \label{eq:nonlinear}
\hat v=\frac{Aktv(t)}{3}, \hspace{1cm} \hat\zeta=-\frac{\zeta_1}{k}>0,
\end{equation}

\begin{figure}
\includegraphics[width=\linewidth]{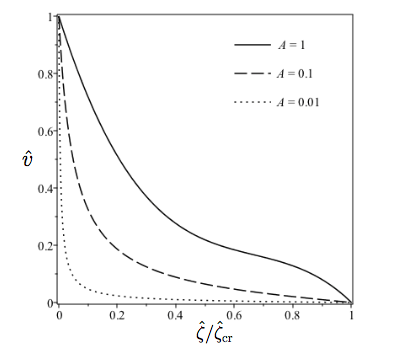}
\caption{Bubble tip velocity as a function of curvature for various Atwood numbers}
\label{fig:tipvelocity}
\end{figure}

We may likewise scale the harmonics:
\[\hat\Phi_{mn}=\frac{Akt\Phi_{mn}(t)}{3}, \hspace{1cm} \hat{\tilde\Phi}_{mn}=\frac{Akt\tilde\Phi_{mn}(t)}{3}.\]
For $\hat\zeta$ small, specifically $\hat\zeta\ll\frac{\sqrt{5}}{8}$,
\[\hat v\ \approx\ 1-\frac{10A^2+16}{3A}\hat\zeta.\]
For bubbles, $\zeta<0$ and $v>0$ in which case solutions will exist for $\hat\zeta\in(0, \hat\zeta_{\rm cr})$ where
$\hat\zeta_{\rm cr}=\frac{3}{8}$ with corresponding $\zeta_{\rm cr}=-\frac{3}{8}k$. Fig. \ref{fig:tipvelocity} shows the bubble tip
velocity as a function of the bubble curvature. In the limits $A\to0^+$ and $A\to1^-$, the velocities become, respectively,
\[\hat v_{A=0}(t)=\frac{9-64\hat\zeta^2}{16kt\hat\zeta}, \quad
\hat v_{A=1}(t)=\frac{(3-8\hat\zeta)(16\hat\zeta^2-6\hat\zeta+1)}{kt}.\]

\begin{figure}
\includegraphics[width=\linewidth]{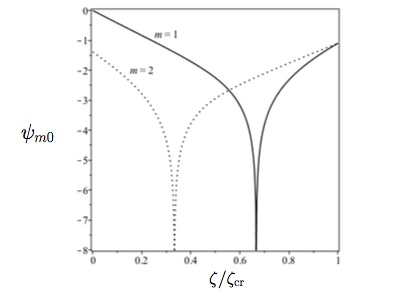}
\caption{$\psi_{m0}$ for the heavy fluid}
\label{fig:heavy}
\end{figure}

\begin{figure}
\includegraphics[width=\linewidth]{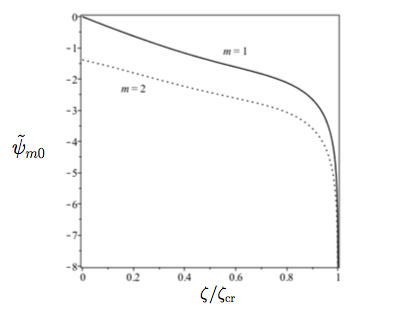}
\caption{$\psi_{m0}$ for the light fluid}
\label{fig:light}
\end{figure}

Figs. \ref{fig:heavy} and \ref{fig:light} are plots of $\psi_{m0}=\ln\left\vert\frac{\Phi_{m0}}{\Phi_{10} {\rm max}}\right\vert$ and 
$\tilde\psi_{m0}=\ln\left\vert\frac{\tilde\Phi_{m0}}{\tilde\Phi_{10} {\rm max}}\right\vert$ as functions of $\zeta/\zeta_{\rm cr}$ for
Atwood number $A=\frac{2}{3}$, and demonstrate that the second Fourier amplitude is much smaller than the first for
$\hat\zeta<\hat\zeta_{\rm cr}$. We note that $\Phi_{10}=\Phi_{20}$ when $\hat\zeta=\frac{5}{24}$. This defines the convergence limit
and we refer to the bubble with this curvature as the `convergence bubble'.

Solutions for $N>1$ can likewise be calculated. The resultant expressions are cumbersome
\cite{anisimov,abarzhireview,dynamicsreview,garabedian,inogamov,abarzhisteady,supernovae} and not given here.
Similarly to these cited works, the solutions converge for increasing $N$ and in each case the lowest order harmonics are dominant

\subsection{The effect of shear}

\begin{figure}
\includegraphics[width=\linewidth]{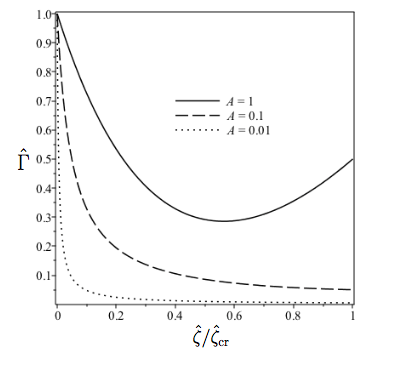}
\caption{Shear as a function of curvature for various Atwood numbers}
\label{fig:shear}
\end{figure}

The multiplicity of these nonlinear asymptotic solutions is also due to the presence of shear at the interface,
as suggested by the boundary conditions Eq. (\ref{eq:momentshear}). We define shear function $\Gamma$
to be the spatial derivative of the jump in the tangential velocity across the interface. We find that in the vicinity of
the bubble tip it is $\Gamma=\tilde M_1-M_1$. Specifically,
\begin{equation} \label{eq:interfaceshear}
\hat\Gamma=\frac{9\hat v}{9-64\hat\zeta^2}=\frac{3A(128A\hat\zeta^3-10A\hat\zeta+3)}{64A\hat\zeta^2+9A+48\hat\zeta},
\end{equation}
where
\[\hat\Gamma(t)=\frac{At}{2}\Gamma.\]
For $\hat\zeta$ small, specifically $\hat\zeta\ll\frac{\sqrt{5}}{8}$,
\[\hat\Gamma\ \approx\ 1-\frac{10A^2+16}{3A}\hat\zeta\]
Fig. \ref{fig:shear} shows the interface shear function versus the bubble curvature. For Atwood numbers exceeding
$A^*=\frac{2}{9}$ the shear function is concave upwards, and for values $A<A^*$ it is monotone decreasing. When $A=1$,
the shear function achieves its minimum value of $\hat\Gamma=\frac{6\sqrt{22}-27}{4}$ at
$\frac{\hat\zeta}{\zeta_{\rm cr}}=\frac{\sqrt{22}-3}{3}$ and the corresponding velocity is $\hat v=\frac{231-49\sqrt{22}}{18}$.

For Atwood numbers $A>A^*$, the shear function attains a minimum value $\hat\Gamma_{\rm min}$ at some curvature value
$\hat\zeta_{\rm min}$ and consequentially, there are two branches of solutions. For small curvatures $\hat\zeta<\hat\zeta_{\rm min}$,
less curved bubbles experience greater shear whereas for large curvatures $\hat\zeta>\hat\zeta_{\rm min}$, less curved bubbles
experience less shear.

\begin{figure}
\includegraphics[width=\linewidth]{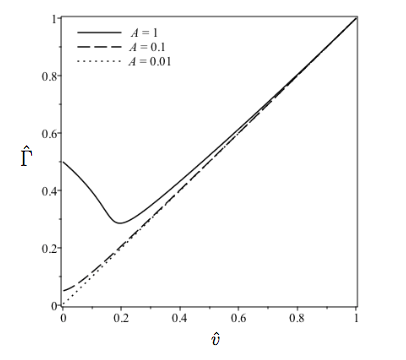}
\caption{Bubble tip velocity as a function of shear for various Atwood numbers}
\label{fig:shearvel}
\end{figure}

Fig. \ref{fig:shearvel} shows the interface shear function versus the bubble tip velocity. For Atwood numbers $A>A^*$, the situation
is as follows: For small curvatures $\hat\zeta<\hat\zeta_{\rm min}$, faster bubbles experience less shear whereas for large curvatures,
$\hat\zeta>\hat\zeta_{\rm min}$, faster bubbles experence more shear.

\subsection{Special solutions}

There are a number of solutions in the family that deserve special attention. These are the fastest bubble, the Taylor bubble,
the convergence bubble, the minimum-shear bubble, and the critical bubble.

\subsubsection{The Atwood bubble}

The fastest member of the family we refer to as the `Atwood bubble' to emphasise its dependence on the Atwood number.
The velocity is $\hat v_{\rm A}=1$, the shear function is $\hat\Gamma_{\rm A}=1$, and the harmonics are
$\hat\Phi_{10{\rm A}}=-\frac{2}{3}\hat v_{\rm a}$, $\hat\Phi_{20{\rm A}}=\frac{1}{6}\hat v_{\rm A}$,
$\hat{\tilde\Phi}_{10{\rm A}}=\frac{2}{3}\hat v_{\rm A}$, $\hat{\tilde\Phi}_{20{\rm A}}=-\frac{1}{6}\hat v_{\rm A}$.
We note that this is in fact the flat bubble.

\subsubsection{The Taylor bubble}

We refer to this bubble as a `Taylor bubble' since its curvature is the same as in the work\cite{davies} except for a difference
in the wavevector value. For the Taylor bubble the curvature, velocity and shear function are
\begin{equation} \label{eq:taylor}
\hat\zeta_{\rm T}=\frac{1}{8}, \hspace{1cm} \hat v_{\rm T}=\frac{4A(3-A)}{3(3+5A)},
\hspace{1cm} \hat\Gamma_{\rm T}=\frac{9}{8}\hat v_T.
\end{equation}
The corresponding Fourier amplitudes are $\hat\Phi_{10{\rm T}}=-\frac{1}{2}\hat v_{\rm T}$, $\hat\Phi_{20{\rm T}}=0$,
$\hat{\tilde\Phi}_{10{\rm T}}=\frac{3}{4}\hat v_{\rm T}$ and $\hat{\tilde\Phi}_{20{\rm T}}=-\frac{1}{4}\hat v_{\rm T}$. Note that
$\hat\Phi_{20{\rm T}}\ne0$ for $N>1$
\cite{anisimov,abarzhireview,dynamicsreview,garabedian,inogamov,abarzhisteady,supernovae}.

\subsubsection{The convergence bubble}

The magnitudes of the Fourier harmonics $\vert\Phi_{10}(t)\vert$ and $\vert\Phi_{20}(t)\vert$ coincide when $\hat\zeta=\frac{5}{24}$.
This defines the convergence limit.  For the convergence bubble the curvature, velocity and shear function are
\begin{equation} \label{eq:conlim}
\hat\zeta_{\rm CL}=\frac{5}{24}, \hspace{.25cm} \hat v_{\rm CL}=\frac{28A(81-25A)}{81(45+53A)},
\hspace{.25cm} \hat \Gamma_{\rm CL}=\frac{81}{56}\hat v_{\rm CL}.
\end{equation}
The corresponding Fourier amplitudes are $\hat\Phi_{10{\rm CL}}=-\frac{1}{4}\hat v_{\rm CL}$,
$\hat\Phi_{20{\rm CL}}=-\frac{1}{4}\hat v_{\rm CL}$, $\hat{\tilde\Phi}_{10{\rm CL}}=\frac{11}{14}\hat v_{\rm CL}$ and
$\hat{\tilde\Phi}_{20{\rm CL}}=-\frac{2}{7}\hat v_{\rm CL}$.

\subsubsection{The minimum-shear bubble}

When $A=1$ the shear function achieves its minimum value $\hat\Gamma=\frac{6\sqrt{22}-27}{4}$ at
$\frac{\hat\zeta}{\hat\zeta_{\rm cr}}=\frac{\sqrt{22}-3}{3}$, and the corresponding velocity is $\hat v=\frac{231-49\sqrt{22}}{18}$.
For $A$ values slightly below $A=1$, the shear function achieves its minimum value at
\[\frac{\hat\zeta}{\hat\zeta_{\rm cr}}=\frac{\sqrt{22}-3}{3}+\frac{1}{5}(1-A).\]
When $A=\frac{2}{9}$, the minimum value for which the shear achieves a minimum in $(0,\zeta_{\rm cr})$, the shear function
achieves its minimum value of $\hat\Gamma=\frac{1}{9}$ at $\frac{\hat\zeta}{\zeta_{\rm cr}}=1$, and the corresponding velocity
is $\hat v=0$. For $A$ values slightly above $A=\frac{2}{9}$, the shear function achieves its minimum value at
\[\frac{\hat\zeta}{\hat\zeta_{\rm cr}}=1-\frac{81}{50}\left(A-\frac{2}{9}\right).\]

\subsubsection{The critical bubble}

For the critical bubble the curvature, velocity and shear function are
\begin{equation} \label{eq:critical}
\hat\zeta_{\rm Cr}=\frac{3}{8}, \hspace{.5cm} \hat v_{\rm Cr}=0, \hspace{.5cm}
\hat\Gamma_{\rm Cr}=\frac{A}{2}.
\end{equation}
The corresponding Fourier amplitudes are $\hat\Phi_{10{\rm Cr}}=\frac{A}{3}$,
$\hat\Phi_{20{\rm Cr}}=-\frac{A}{3}$, $\hat{\tilde\Phi}_{10{\rm Cr}}=0$, $\hat{\tilde\Phi}_{20{\rm Cr}}=0$. Note that
$\hat{\tilde\Phi}_{10{\rm Cr}}\ne0$ and $\hat{\tilde\Phi}_{20{\rm Cr}}\ne0$ for $N>1$.

\subsection{Stability}

\begin{figure}
\includegraphics[width=\linewidth]{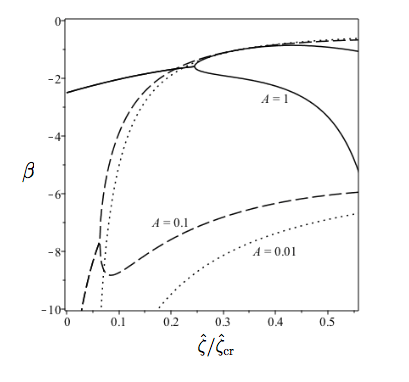}
\caption{Stability profiles for various Atwood numbers}
\label{fig:beta}
\end{figure}

The stability variable $\beta$ satisfies a quadratic equation which does not depend on the value of the acceleration exponent $a$.
The analytical result is too cumbersome to be presented here. Fig \ref{fig:beta} shows the stability function for various Atwood numbers.
We observe that all bubbles up to the convergence limit $\hat\zeta_{\rm CL}/\hat\zeta_{\rm cr}=\frac{5}{9}$ are stable at $N=1$.
The $N>1$ analysis is to be presented elsewhere.

\subsection{Properties of nonlinear RM dynamics}

\subsubsection{Multiscale character of RM dynamics}

The multi-scale character of the dynamics can be understood by viewing the RM coherent structure as a standing wave with growing
amplitude \cite{landau}. The multi-scale character of nonlinear RMI is consistent with the existence of an amplitude scale in early-time
shock-driven RMI, at which the maximum initial growth-rate of RMI is achieved \cite{dell2017}. 

The Atwood (flat) bubble is the fastest stable solution and is hence the physically significant solution. This solution has the
(quasi) invariant value
\[\frac{tv_{\rm F}^2}{\left(\frac{dv_{\rm F}}{d\zeta}\right)_{\zeta=0}}=\frac{9}{16+10A^2}.\]
This (quasi) invariance implies that nonlinear RM dynamics is multi-scale, with two macroscopic length scales contributing. These
being the wavelength and the amplitude \cite{anisimov,abarzhireview,dynamicsreview}.

\subsubsection{Interfacial character of RM dynamics}

\begin{figure}
\includegraphics[width=1\linewidth]{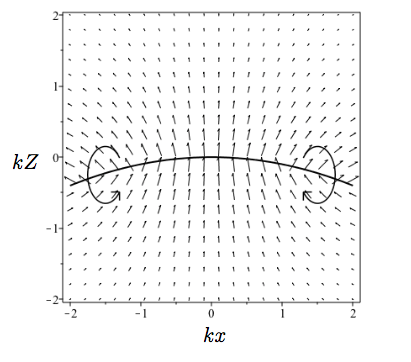}
\caption{Qualitative velocity field of the Atwood bubble in the plane $y=0$ in the laboratory reference frame, $Z=z-z_0(t)$.
The curved arrows indicate that near the tip of the bubble the vortical structures `rotate' from the heavy to the light fluid.}
\label{fig:velocityfield}
\end{figure}

By accurately accounting for the interplay of harmonics and by systematically connecting the interfacial velocity and shear
for a broad range of acceleration parameters, we have found that RM dynamics is essentially interfacial: It has intense
fluid motion in the vicinity of the interface and effectively no motion away from the interface. The velocity is potential in the
bulk of each fluid. Shear-driven vortical structures may appear at the interface. Fig. \ref{fig:velocityfield} shows the qualitative
velocity field in the laboratory reference frame in the $(x,z)$-plane of the $\hat\zeta=1/10$ bubble for Atwood number
$A=2/3$ at time $t=1$, for any $a<-2$. Shear-driven vortical structures may appear at the interface due to discontinuity
of the tangential component of velocity. Near the tip of the bubble the vortical structures `rotate' from the heavy to the light fluid.
This velocity pattern is observed in experiments and simulations, demonstrating qualitative agreement with our results
\cite{meshkov,meshkov2013,robey,kadau,glimm,youngs,supernovae}.

\section{Discussion}

Our theory finds that in RMI with variable acceleration, nonlinear bubbles decelerate and flatten. This behaviour is observed in
experiments and simulations, in agreement with our results
\cite{abarzhireview,meshkovbook,dynamicsreview,stanic,dell2015,dell2017,psa,meshkov2013,supernovae}. According to our  theory,
in nonlinear RMI with variable acceleration, flattened bubbles move more quickly and decelerate more rapidly when compared
to curved bubbles because since the bubble velocity decays with time as $~C/kt$, the deceleration is $~-C/kt^2$.
Moreover, according to our results, RM bubbles move more quickly and have larger interfacial shear for fluids with similar densities
than for fluids with very different densities \cite{meshkovbook,dynamicsreview,meshkov2013,bhowmick2016,thornber,hbia}. This
result has a clear interpretation: for fluids with similar densities, shear-driven interfacial vortical structures are more intense, leading
to stronger energy dissipation, stronger deceleration and thus, to larger bubble velocity when compared to fluids with very different
densities.

Our analysis is focused on large-scale dynamics, presuming that interfacial vortical structures are small. This assumption is
applicable for fluids of very different densities and with a finite density ratio. For fluids with very similar densities $A\to0^+$ other
approaches should be employed \cite{anisimov,abarzhireview,dynamicsreview,abarzhisteady,supernovae}. While for fluids with
very similar densities our theory is no longer applicable, the singular nature of the velocity of the fastest stable bubble for $A\to0^+$
indicates that for $A\to0^+$ and $t/\tau\to\infty$ the bubble velocity may decay more quickly than inverse time
\cite{abarzhireview,meshkovbook,dynamicsreview,meshkov2013,anisimov,ang}.

According to our results, for variable acceleration with $a<-2$, RM dynamics depends on the initial conditions and is independent
of the acceleration. Hence, one can scrupulously study the effect of initial conditions on RM dynamics by analyzing properties of the
unstable interface for various accelerations \cite{arnett,zeldovichraizer,haan,peters,rana,buehler,supernovae}. Note that accurate
quantification of nonlinear RMI in observations may be a challenge because the interface velocity is usually of the order of
$~0.1\%$ of the largest velocity scale in the post-shock fluid system, and because the interface velocity is a power-law function of
time, which requires a substantial span of temporal and spatial scales for accurate diagnostics
\cite{dynamicsreview,stanic,dell2015,dell2017,psa,meshkov2013}.

In addition to determining the interface velocity, we have elaborated theory benchmarks which have not been discussed before.
These are the fields of velocity and pressure, interface morphology and bubble curvature, interfacial shear and its link to the bubble
velocity and curvature, and spectral properties of the velocity and pressure. By diagnosing the dependence of these quantities on the
density ratios, flow symmetries, initial conditions, and accelerations, by identifying their universal properties, and by accurately
measuring departures of data in real fluids from theoretical solutions in ideal fluids, one can further advance knowledge of RM dynamics
in realistic environments, better understand RM relevant  processes in nature and technology, and improve methods of numerical
modeling and experimental diagnostics of RM dynamics in fluids, plasmas, and materials.

To conclude, we have considered the long-standing problem of RMI with variable acceleration by applying group theory. We have
directly linked the interface velocity, morphology and shear, revealed the interfacial and multi-scale character of RM dynamics,
achieved good agreement with available observations, and elaborated new theory benchmarks for future experiments and
simulations.

\section{Acknowledgements}

The authors thank the University of Western Australia (AUS) and the National Science Foundation (USA).

\section{References}

\bibliographystyle{apsrev4-1}
\bibliography{bb}

\begin{thebibliography}{41}%
\makeatletter
\providecommand \@ifxundefined [1]{%
 \@ifx{#1\undefined}
}%
\providecommand \@ifnum [1]{%
 \ifnum #1\expandafter \@firstoftwo
 \else \expandafter \@secondoftwo
 \fi
}%
\providecommand \@ifx [1]{%
 \ifx #1\expandafter \@firstoftwo
 \else \expandafter \@secondoftwo
 \fi
}%
\providecommand \natexlab [1]{#1}%
\providecommand \enquote  [1]{``#1''}%
\providecommand \bibnamefont  [1]{#1}%
\providecommand \bibfnamefont [1]{#1}%
\providecommand \citenamefont [1]{#1}%
\providecommand \href@noop [0]{\@secondoftwo}%
\providecommand \href [0]{\begingroup \@sanitize@url \@href}%
\providecommand \@href[1]{\@@startlink{#1}\@@href}%
\providecommand \@@href[1]{\endgroup#1\@@endlink}%
\providecommand \@sanitize@url [0]{\catcode `\\12\catcode `\$12\catcode
  `\&12\catcode `\#12\catcode `\^12\catcode `\_12\catcode `\%12\relax}%
\providecommand \@@startlink[1]{}%
\providecommand \@@endlink[0]{}%
\providecommand \url  [0]{\begingroup\@sanitize@url \@url }%
\providecommand \@url [1]{\endgroup\@href {#1}{\urlprefix }}%
\providecommand \urlprefix  [0]{URL }%
\providecommand \Eprint [0]{\href }%
\providecommand \doibase [0]{http://dx.doi.org/}%
\providecommand \selectlanguage [0]{\@gobble}%
\providecommand \bibinfo  [0]{\@secondoftwo}%
\providecommand \bibfield  [0]{\@secondoftwo}%
\providecommand \translation [1]{[#1]}%
\providecommand \BibitemOpen [0]{}%
\providecommand \bibitemStop [0]{}%
\providecommand \bibitemNoStop [0]{.\EOS\space}%
\providecommand \EOS [0]{\spacefactor3000\relax}%
\providecommand \BibitemShut  [1]{\csname bibitem#1\endcsname}%
\let\auto@bib@innerbib\@empty
%</preamble>
\bibitem [{\citenamefont {Richtmyer}(1960)}]{richtmyer}%
  \BibitemOpen
  \bibfield  {author} {\bibinfo {author} {\bibfnamefont {R.~D.}\ \bibnamefont
  {Richtmyer}},\ }\href@noop {} {\bibfield  {journal} {\bibinfo  {journal}
  {Commun Pure Appl Math}\ }\textbf {\bibinfo {volume} {13}},\ \bibinfo {pages}
  {297} (\bibinfo {year} {1960})}\BibitemShut {NoStop}%
\bibitem [{\citenamefont {Meshkov}(1969)}]{meshkov}%
  \BibitemOpen
  \bibfield  {author} {\bibinfo {author} {\bibfnamefont {E.~E.}\ \bibnamefont
  {Meshkov}},\ }\href@noop {} {\bibfield  {journal} {\bibinfo  {journal} {Sov
  Fluid Dyn}\ }\textbf {\bibinfo {volume} {4}},\ \bibinfo {pages} {101}
  (\bibinfo {year} {1969})}\BibitemShut {NoStop}%
\bibitem [{\citenamefont {Rayleigh}(1883)}]{rayleigh}%
  \BibitemOpen
  \bibfield  {author} {\bibinfo {author} {\bibfnamefont {L.}~\bibnamefont
  {Rayleigh}},\ }\href@noop {} {\bibfield  {journal} {\bibinfo  {journal} {Proc
  London Math Soc}\ }\textbf {\bibinfo {volume} {14}},\ \bibinfo {pages} {170}
  (\bibinfo {year} {1883})}\BibitemShut {NoStop}%
\bibitem [{\citenamefont {Davies}\ and\ \citenamefont {Taylor}(1950)}]{davies}%
  \BibitemOpen
  \bibfield  {author} {\bibinfo {author} {\bibfnamefont {R.~M.}\ \bibnamefont
  {Davies}}\ and\ \bibinfo {author} {\bibfnamefont {G.~I.}\ \bibnamefont
  {Taylor}},\ }\href@noop {} {\bibfield  {journal} {\bibinfo  {journal} {Proc R
  Soc A}\ }\textbf {\bibinfo {volume} {200}},\ \bibinfo {pages} {375} (\bibinfo
  {year} {1950})}\BibitemShut {NoStop}%
\bibitem [{\citenamefont {Abarzhi}(2010)}]{abarzhireview}%
  \BibitemOpen
  \bibfield  {author} {\bibinfo {author} {\bibfnamefont {S.~I.}\ \bibnamefont
  {Abarzhi}},\ }\href@noop {} {\bibfield  {journal} {\bibinfo  {journal} {Phil
  Trans R Soc A}\ }\textbf {\bibinfo {volume} {368}},\ \bibinfo {pages} {1809}
  (\bibinfo {year} {2010})}\BibitemShut {NoStop}%
\bibitem [{\citenamefont {Meshkov}(2006)}]{meshkovbook}%
  \BibitemOpen
  \bibfield  {author} {\bibinfo {author} {\bibfnamefont {E.~E.}\ \bibnamefont
  {Meshkov}},\ }\href@noop {} {\emph {\bibinfo {title} {Studies of hydrodynamic
  instabilities in laboratory experiments}}}\ (\bibinfo  {publisher} {Sarov},\
  \bibinfo {year} {2006})\BibitemShut {NoStop}%
\bibitem [{\citenamefont {Abarzhi}\ \emph {et~al.}(2013)\citenamefont
  {Abarzhi}, \citenamefont {Gauthier},\ and\ \citenamefont
  {Sreenivasan}}]{turbulentmixing}%
  \BibitemOpen
  \bibfield  {author} {\bibinfo {author} {\bibfnamefont {S.~I.}\ \bibnamefont
  {Abarzhi}}, \bibinfo {author} {\bibfnamefont {S.}~\bibnamefont {Gauthier}}, \
  and\ \bibinfo {author} {\bibfnamefont {K.~R.}\ \bibnamefont {Sreenivasan}},\
  }\href@noop {} {\emph {\bibinfo {title} {Turbulent mixing and beyond:
  non-equilibrium processes from atomistic to astrophysical scales. I \& II}}}\
  (\bibinfo  {publisher} {Royal Society Publishing},\ \bibinfo {year}
  {2013})\BibitemShut {NoStop}%
\bibitem [{\citenamefont {Arnett}(1996)}]{arnett}%
  \BibitemOpen
  \bibfield  {author} {\bibinfo {author} {\bibfnamefont {D.}~\bibnamefont
  {Arnett}},\ }\href@noop {} {\emph {\bibinfo {title} {Supernovae and
  Nucleosynthesis: An Investigation of the History of Matter, from the Big Bang
  to the Present}}}\ (\bibinfo  {publisher} {Princeton University Press},\
  \bibinfo {year} {1996})\BibitemShut {NoStop}%
\bibitem [{\citenamefont {Zeldovich}\ and\ \citenamefont
  {Raizer}(2002)}]{zeldovichraizer}%
  \BibitemOpen
  \bibfield  {author} {\bibinfo {author} {\bibfnamefont {Y.~B.}\ \bibnamefont
  {Zeldovich}}\ and\ \bibinfo {author} {\bibfnamefont {Y.~P.}\ \bibnamefont
  {Raizer}},\ }\href@noop {} {\emph {\bibinfo {title} {Physics of shock waves
  and high-temperature hydrodynamic phenomena}}}\ (\bibinfo  {publisher} {Dover
  New York},\ \bibinfo {year} {2002})\BibitemShut {NoStop}%
\bibitem [{\citenamefont {Haan}(2011)}]{haan}%
  \BibitemOpen
  \bibfield  {author} {\bibinfo {author} {\bibfnamefont {S.~W.}\ \bibnamefont
  {Haan}},\ }\href@noop {} {\bibfield  {journal} {\bibinfo  {journal} {Phys.
  Plasmas}\ }\textbf {\bibinfo {volume} {18}},\ \bibinfo {pages} {051001}
  (\bibinfo {year} {2011})}\BibitemShut {NoStop}%
\bibitem [{\citenamefont {Peters}(2000)}]{peters}%
  \BibitemOpen
  \bibfield  {author} {\bibinfo {author} {\bibfnamefont {N.}~\bibnamefont
  {Peters}},\ }\href@noop {} {\emph {\bibinfo {title} {Turbulent Combustion}}}\
  (\bibinfo  {publisher} {Cambridge University Press},\ \bibinfo {year}
  {2000})\BibitemShut {NoStop}%
\bibitem [{\citenamefont {Rana}\ and\ \citenamefont {Herrmann}(2011)}]{rana}%
  \BibitemOpen
  \bibfield  {author} {\bibinfo {author} {\bibfnamefont {S.}~\bibnamefont
  {Rana}}\ and\ \bibinfo {author} {\bibfnamefont {M.}~\bibnamefont
  {Herrmann}},\ }\href@noop {} {\bibfield  {journal} {\bibinfo  {journal} {Phys
  Fluids}\ }\textbf {\bibinfo {volume} {23}},\ \bibinfo {pages} {091109}
  (\bibinfo {year} {2011})}\BibitemShut {NoStop}%
\bibitem [{\citenamefont {Buehler}\ \emph {et~al.}(2007)\citenamefont
  {Buehler}, \citenamefont {Tang}, \citenamefont {van Duin},\ and\
  \citenamefont {Goddard}}]{buehler}%
  \BibitemOpen
  \bibfield  {author} {\bibinfo {author} {\bibfnamefont {M.~J.}\ \bibnamefont
  {Buehler}}, \bibinfo {author} {\bibfnamefont {H.}~\bibnamefont {Tang}},
  \bibinfo {author} {\bibfnamefont {A.~C.~T.}\ \bibnamefont {van Duin}}, \ and\
  \bibinfo {author} {\bibfnamefont {W.~A.}\ \bibnamefont {Goddard}},\
  }\href@noop {} {\bibfield  {journal} {\bibinfo  {journal} {Phys Rev Lett}\
  }\textbf {\bibinfo {volume} {99}},\ \bibinfo {pages} {165502} (\bibinfo
  {year} {2007})}\BibitemShut {NoStop}%
\bibitem [{\citenamefont {Abarzhi}\ \emph {et~al.}(2018)\citenamefont
  {Abarzhi}, \citenamefont {Bhowmick}, \citenamefont {Naveh}, \citenamefont
  {Pandian}, \citenamefont {Swisher}, \citenamefont {Stellingwerf},\ and\
  \citenamefont {Arnett}}]{supernovae}%
  \BibitemOpen
  \bibfield  {author} {\bibinfo {author} {\bibfnamefont {S.~I.}\ \bibnamefont
  {Abarzhi}}, \bibinfo {author} {\bibfnamefont {A.}~\bibnamefont {Bhowmick}},
  \bibinfo {author} {\bibfnamefont {A.}~\bibnamefont {Naveh}}, \bibinfo
  {author} {\bibfnamefont {A.}~\bibnamefont {Pandian}}, \bibinfo {author}
  {\bibfnamefont {N.}~\bibnamefont {Swisher}}, \bibinfo {author} {\bibfnamefont
  {R.}~\bibnamefont {Stellingwerf}}, \ and\ \bibinfo {author} {\bibfnamefont
  {W.}~\bibnamefont {Arnett}},\ }\href@noop {} {\bibfield  {journal} {\bibinfo
  {journal} {Proc Natl Acad Sci USA}\ ,\ \bibinfo {pages} {201714502}}
  (\bibinfo {year} {2018})}\BibitemShut {NoStop}%
\bibitem [{\citenamefont {Abarzhi}(2008)}]{dynamicsreview}%
  \BibitemOpen
  \bibfield  {author} {\bibinfo {author} {\bibfnamefont {S.~I.}\ \bibnamefont
  {Abarzhi}},\ }\href@noop {} {\bibfield  {journal} {\bibinfo  {journal}
  {Physica Scripta}\ }\textbf {\bibinfo {volume} {2008 (T132)}},\ \bibinfo
  {pages} {014012} (\bibinfo {year} {2008})}\BibitemShut {NoStop}%
\bibitem [{\citenamefont {Stanic}\ \emph {et~al.}(2012)\citenamefont {Stanic},
  \citenamefont {Stellingwerf}, \citenamefont {Cassibry},\ and\ \citenamefont
  {Abarzhi}}]{stanic}%
  \BibitemOpen
  \bibfield  {author} {\bibinfo {author} {\bibfnamefont {M.}~\bibnamefont
  {Stanic}}, \bibinfo {author} {\bibfnamefont {R.~F.}\ \bibnamefont
  {Stellingwerf}}, \bibinfo {author} {\bibfnamefont {J.~T.}\ \bibnamefont
  {Cassibry}}, \ and\ \bibinfo {author} {\bibfnamefont {S.~I.}\ \bibnamefont
  {Abarzhi}},\ }\href@noop {} {\bibfield  {journal} {\bibinfo  {journal} {Phys
  Plasmas}\ }\textbf {\bibinfo {volume} {19}},\ \bibinfo {pages} {082706}
  (\bibinfo {year} {2012})}\BibitemShut {NoStop}%
\bibitem [{\citenamefont {Dell}\ \emph {et~al.}(2015)\citenamefont {Dell},
  \citenamefont {Stellingwerf},\ and\ \citenamefont {Abarzhi}}]{dell2015}%
  \BibitemOpen
  \bibfield  {author} {\bibinfo {author} {\bibfnamefont {Z.}~\bibnamefont
  {Dell}}, \bibinfo {author} {\bibfnamefont {R.~F.}\ \bibnamefont
  {Stellingwerf}}, \ and\ \bibinfo {author} {\bibfnamefont {S.~I.}\
  \bibnamefont {Abarzhi}},\ }\href@noop {} {\bibfield  {journal} {\bibinfo
  {journal} {Phys Plasmas}\ }\textbf {\bibinfo {volume} {22}},\ \bibinfo
  {pages} {092711} (\bibinfo {year} {2015})}\BibitemShut {NoStop}%
\bibitem [{\citenamefont {Dell}\ \emph {et~al.}(2017)\citenamefont {Dell} \emph
  {et~al.}}]{dell2017}%
  \BibitemOpen
  \bibfield  {author} {\bibinfo {author} {\bibfnamefont {Z.}~\bibnamefont
  {Dell}} \emph {et~al.},\ }\href@noop {} {\bibfield  {journal} {\bibinfo
  {journal} {Phys Plasmas}\ }\textbf {\bibinfo {volume} {24}},\ \bibinfo
  {pages} {090702} (\bibinfo {year} {2017})}\BibitemShut {NoStop}%
\bibitem [{\citenamefont {Pandian}\ \emph {et~al.}()\citenamefont {Pandian},
  \citenamefont {Stellingwerf},\ and\ \citenamefont {Abarzhi}}]{psa}%
  \BibitemOpen
  \bibfield  {author} {\bibinfo {author} {\bibfnamefont {A.}~\bibnamefont
  {Pandian}}, \bibinfo {author} {\bibfnamefont {R.~F.}\ \bibnamefont
  {Stellingwerf}}, \ and\ \bibinfo {author} {\bibfnamefont {S.~I.}\
  \bibnamefont {Abarzhi}},\ }\href@noop {} {\ }\BibitemShut {NoStop}%
\bibitem [{\citenamefont {Meshkov}(2013)}]{meshkov2013}%
  \BibitemOpen
  \bibfield  {author} {\bibinfo {author} {\bibfnamefont {E.~E.}\ \bibnamefont
  {Meshkov}},\ }\href@noop {} {\bibfield  {journal} {\bibinfo  {journal} {Phil
  Trans R Soc A}\ }\textbf {\bibinfo {volume} {371}},\ \bibinfo {pages}
  {20120288} (\bibinfo {year} {2013})}\BibitemShut {NoStop}%
\bibitem [{\citenamefont {Robey}\ \emph {et~al.}(2003)\citenamefont {Robey},
  \citenamefont {Zhou}, \citenamefont {Buckingham}, \citenamefont {Keiter},
  \citenamefont {Remington},\ and\ \citenamefont {Drake}}]{robey}%
  \BibitemOpen
  \bibfield  {author} {\bibinfo {author} {\bibfnamefont {H.~F.}\ \bibnamefont
  {Robey}}, \bibinfo {author} {\bibfnamefont {Y.}~\bibnamefont {Zhou}},
  \bibinfo {author} {\bibfnamefont {A.~C.}\ \bibnamefont {Buckingham}},
  \bibinfo {author} {\bibfnamefont {P.}~\bibnamefont {Keiter}}, \bibinfo
  {author} {\bibfnamefont {B.~A.}\ \bibnamefont {Remington}}, \ and\ \bibinfo
  {author} {\bibfnamefont {R.~P.}\ \bibnamefont {Drake}},\ }\href@noop {}
  {\bibfield  {journal} {\bibinfo  {journal} {Phys. Plasmas}\ }\textbf
  {\bibinfo {volume} {10}},\ \bibinfo {pages} {614} (\bibinfo {year}
  {2003})}\BibitemShut {NoStop}%
\bibitem [{\citenamefont {Lugomer}(2016)}]{lugomer}%
  \BibitemOpen
  \bibfield  {author} {\bibinfo {author} {\bibfnamefont {S.}~\bibnamefont
  {Lugomer}},\ }\href@noop {} {\bibfield  {journal} {\bibinfo  {journal} {Laser
  Part. Beams}\ }\textbf {\bibinfo {volume} {34}},\ \bibinfo {pages} {123}
  (\bibinfo {year} {2016})}\BibitemShut {NoStop}%
\bibitem [{\citenamefont {Swisher}\ \emph {et~al.}(2015)\citenamefont {Swisher}
  \emph {et~al.}}]{swisher}%
  \BibitemOpen
  \bibfield  {author} {\bibinfo {author} {\bibfnamefont {N.}~\bibnamefont
  {Swisher}} \emph {et~al.},\ }\href@noop {} {\bibfield  {journal} {\bibinfo
  {journal} {Phys. Plasmas}\ }\textbf {\bibinfo {volume} {22}},\ \bibinfo
  {pages} {102707} (\bibinfo {year} {2015})}\BibitemShut {NoStop}%
\bibitem [{\citenamefont {Orlov}\ \emph {et~al.}(2010)\citenamefont {Orlov},
  \citenamefont {Abarzhi}, \citenamefont {Oh}, \citenamefont {Barbastathis},\
  and\ \citenamefont {Sreenivasan}}]{orlov}%
  \BibitemOpen
  \bibfield  {author} {\bibinfo {author} {\bibfnamefont {S.~S.}\ \bibnamefont
  {Orlov}}, \bibinfo {author} {\bibfnamefont {S.~I.}\ \bibnamefont {Abarzhi}},
  \bibinfo {author} {\bibfnamefont {S.-B.}\ \bibnamefont {Oh}}, \bibinfo
  {author} {\bibfnamefont {G.}~\bibnamefont {Barbastathis}}, \ and\ \bibinfo
  {author} {\bibfnamefont {K.~R.}\ \bibnamefont {Sreenivasan}},\ }\href@noop {}
  {\bibfield  {journal} {\bibinfo  {journal} {Phil Trans R Soc A}\ }\textbf
  {\bibinfo {volume} {368}},\ \bibinfo {pages} {1705} (\bibinfo {year}
  {2010})}\BibitemShut {NoStop}%
\bibitem [{\citenamefont {Anisimov}\ \emph {et~al.}(2013)\citenamefont
  {Anisimov}, \citenamefont {Drake}, \citenamefont {Gauthier}, \citenamefont
  {Meshkov},\ and\ \citenamefont {Abarzhi}}]{anisimov}%
  \BibitemOpen
  \bibfield  {author} {\bibinfo {author} {\bibfnamefont {S.~I.}\ \bibnamefont
  {Anisimov}}, \bibinfo {author} {\bibfnamefont {R.~P.}\ \bibnamefont {Drake}},
  \bibinfo {author} {\bibfnamefont {S.}~\bibnamefont {Gauthier}}, \bibinfo
  {author} {\bibfnamefont {E.~E.}\ \bibnamefont {Meshkov}}, \ and\ \bibinfo
  {author} {\bibfnamefont {S.~I.}\ \bibnamefont {Abarzhi}},\ }\href@noop {}
  {\bibfield  {journal} {\bibinfo  {journal} {Phil Trans R Soc A}\ }\textbf
  {\bibinfo {volume} {371}},\ \bibinfo {pages} {20130266} (\bibinfo {year}
  {2013})}\BibitemShut {NoStop}%
\bibitem [{\citenamefont {Kull}(1991)}]{kull}%
  \BibitemOpen
  \bibfield  {author} {\bibinfo {author} {\bibfnamefont {H.~J.}\ \bibnamefont
  {Kull}},\ }\href@noop {} {\bibfield  {journal} {\bibinfo  {journal} {Phys
  Rep}\ }\textbf {\bibinfo {volume} {206}},\ \bibinfo {pages} {197} (\bibinfo
  {year} {1991})}\BibitemShut {NoStop}%
\bibitem [{\citenamefont {Abarzhi}(1998)}]{abarzhisteady}%
  \BibitemOpen
  \bibfield  {author} {\bibinfo {author} {\bibfnamefont {S.~I.}\ \bibnamefont
  {Abarzhi}},\ }\href@noop {} {\bibfield  {journal} {\bibinfo  {journal} {Phys
  Rev Lett}\ }\textbf {\bibinfo {volume} {81}},\ \bibinfo {pages} {337}
  (\bibinfo {year} {1998})}\BibitemShut {NoStop}%
\bibitem [{\citenamefont {Nishihara}\ \emph {et~al.}(2010)\citenamefont
  {Nishihara}, \citenamefont {Wouchuk}, \citenamefont {Matsuoka}, \citenamefont
  {Ishizaki},\ and\ \citenamefont {Zhakhovsky}}]{nishihara}%
  \BibitemOpen
  \bibfield  {author} {\bibinfo {author} {\bibfnamefont {K.}~\bibnamefont
  {Nishihara}}, \bibinfo {author} {\bibfnamefont {J.~G.}\ \bibnamefont
  {Wouchuk}}, \bibinfo {author} {\bibfnamefont {C.}~\bibnamefont {Matsuoka}},
  \bibinfo {author} {\bibfnamefont {R.}~\bibnamefont {Ishizaki}}, \ and\
  \bibinfo {author} {\bibfnamefont {V.~V.}\ \bibnamefont {Zhakhovsky}},\
  }\href@noop {} {\bibfield  {journal} {\bibinfo  {journal} {Phil Trans R Soc
  A}\ }\textbf {\bibinfo {volume} {368}},\ \bibinfo {pages} {1769} (\bibinfo
  {year} {2010})}\BibitemShut {NoStop}%
\bibitem [{\citenamefont {Bhowmick}\ and\ \citenamefont
  {Abarzhi}(2016)}]{bhowmick2016}%
  \BibitemOpen
  \bibfield  {author} {\bibinfo {author} {\bibfnamefont {A.~K.}\ \bibnamefont
  {Bhowmick}}\ and\ \bibinfo {author} {\bibfnamefont {S.~I.}\ \bibnamefont
  {Abarzhi}},\ }\href@noop {} {\bibfield  {journal} {\bibinfo  {journal} {Phys
  Plasmas}\ }\textbf {\bibinfo {volume} {23}},\ \bibinfo {pages} {112702}
  (\bibinfo {year} {2016})}\BibitemShut {NoStop}%
\bibitem [{\citenamefont {Gauthier}\ and\ \citenamefont
  {Creurer}(2010)}]{gauthier}%
  \BibitemOpen
  \bibfield  {author} {\bibinfo {author} {\bibfnamefont {S.}~\bibnamefont
  {Gauthier}}\ and\ \bibinfo {author} {\bibfnamefont {B.~L.}\ \bibnamefont
  {Creurer}},\ }\href@noop {} {\bibfield  {journal} {\bibinfo  {journal} {Phil
  Trans R Soc A}\ }\textbf {\bibinfo {volume} {368}},\ \bibinfo {pages}
  {368:1681} (\bibinfo {year} {2010})}\BibitemShut {NoStop}%
\bibitem [{\citenamefont {Kadau}\ \emph {et~al.}(2010)\citenamefont {Kadau},
  \citenamefont {Barber}, \citenamefont {Germann}, \citenamefont {Holian},\
  and\ \citenamefont {Alder}}]{kadau}%
  \BibitemOpen
  \bibfield  {author} {\bibinfo {author} {\bibfnamefont {K.}~\bibnamefont
  {Kadau}}, \bibinfo {author} {\bibfnamefont {J.~L.}\ \bibnamefont {Barber}},
  \bibinfo {author} {\bibfnamefont {T.~C.}\ \bibnamefont {Germann}}, \bibinfo
  {author} {\bibfnamefont {B.~L.}\ \bibnamefont {Holian}}, \ and\ \bibinfo
  {author} {\bibfnamefont {B.~J.}\ \bibnamefont {Alder}},\ }\href@noop {}
  {\bibfield  {journal} {\bibinfo  {journal} {Phil Trans R Soc A}\ }\textbf
  {\bibinfo {volume} {368}},\ \bibinfo {pages} {1547} (\bibinfo {year}
  {2010})}\BibitemShut {NoStop}%
\bibitem [{\citenamefont {Glimm}\ \emph {et~al.}(2013)\citenamefont {Glimm},
  \citenamefont {Sharp}, \citenamefont {Kaman},\ and\ \citenamefont
  {Lim}}]{glimm}%
  \BibitemOpen
  \bibfield  {author} {\bibinfo {author} {\bibfnamefont {J.}~\bibnamefont
  {Glimm}}, \bibinfo {author} {\bibfnamefont {D.~H.}\ \bibnamefont {Sharp}},
  \bibinfo {author} {\bibfnamefont {T.}~\bibnamefont {Kaman}}, \ and\ \bibinfo
  {author} {\bibfnamefont {H.}~\bibnamefont {Lim}},\ }\href@noop {} {\bibfield
  {journal} {\bibinfo  {journal} {Phil Trans R Soc A}\ }\textbf {\bibinfo
  {volume} {371}},\ \bibinfo {pages} {20120183} (\bibinfo {year}
  {2013})}\BibitemShut {NoStop}%
\bibitem [{\citenamefont {Youngs}(2013)}]{youngs}%
  \BibitemOpen
  \bibfield  {author} {\bibinfo {author} {\bibfnamefont {D.~L.}\ \bibnamefont
  {Youngs}},\ }\href@noop {} {\bibfield  {journal} {\bibinfo  {journal} {Phil
  Trans R Soc A}\ }\textbf {\bibinfo {volume} {371}},\ \bibinfo {pages}
  {20120173} (\bibinfo {year} {2013})}\BibitemShut {NoStop}%
\bibitem [{\citenamefont {Thornber}\ \emph {et~al.}(2017)\citenamefont
  {Thornber} \emph {et~al.}}]{thornber}%
  \BibitemOpen
  \bibfield  {author} {\bibinfo {author} {\bibfnamefont {B.}~\bibnamefont
  {Thornber}} \emph {et~al.},\ }\href@noop {} {\bibfield  {journal} {\bibinfo
  {journal} {Phys Fluids}\ }\textbf {\bibinfo {volume} {29}},\ \bibinfo {pages}
  {105107} (\bibinfo {year} {2017})}\BibitemShut {NoStop}%
\bibitem [{\citenamefont {Landau}\ and\ \citenamefont
  {Lifshitz}(1987)}]{landau}%
  \BibitemOpen
  \bibfield  {author} {\bibinfo {author} {\bibfnamefont {L.~D.}\ \bibnamefont
  {Landau}}\ and\ \bibinfo {author} {\bibfnamefont {E.~M.}\ \bibnamefont
  {Lifshitz}},\ }\href@noop {} {\emph {\bibinfo {title} {Course of Theoretical
  Physics}}}\ (\bibinfo  {publisher} {Pergamon Press, New York},\ \bibinfo
  {year} {1987})\BibitemShut {NoStop}%
\bibitem [{\citenamefont {Sedov}(1993)}]{sedov}%
  \BibitemOpen
  \bibfield  {author} {\bibinfo {author} {\bibfnamefont {L.~I.}\ \bibnamefont
  {Sedov}},\ }\href@noop {} {\emph {\bibinfo {title} {Similarity and
  dimensional methods in mechanics}}}\ (\bibinfo  {publisher} {CRC Press},\
  \bibinfo {year} {1993})\BibitemShut {NoStop}%
\bibitem [{\citenamefont {Hill}\ \emph
  {et~al.}(2019{\natexlab{a}})\citenamefont {Hill}, \citenamefont {Bhowmich},
  \citenamefont {Ilyin},\ and\ \citenamefont {Abarzhi}}]{hbia}%
  \BibitemOpen
  \bibfield  {author} {\bibinfo {author} {\bibfnamefont {D.~L.}\ \bibnamefont
  {Hill}}, \bibinfo {author} {\bibfnamefont {A.~K.}\ \bibnamefont {Bhowmich}},
  \bibinfo {author} {\bibfnamefont {D.~V.}\ \bibnamefont {Ilyin}}, \ and\
  \bibinfo {author} {\bibfnamefont {S.~I.}\ \bibnamefont {Abarzhi}},\
  }\href@noop {} {\bibfield  {journal} {\bibinfo  {journal} {Phys Rev Fluids}\
  }\textbf {\bibinfo {volume} {4}},\ \bibinfo {pages} {063905} (\bibinfo {year}
  {2019}{\natexlab{a}})}\BibitemShut {NoStop}%
\bibitem [{\citenamefont {Garabedian}(1957)}]{garabedian}%
  \BibitemOpen
  \bibfield  {author} {\bibinfo {author} {\bibfnamefont {P.~R.}\ \bibnamefont
  {Garabedian}},\ }\href@noop {} {\bibfield  {journal} {\bibinfo  {journal}
  {Proc R Soc A}\ }\textbf {\bibinfo {volume} {241}},\ \bibinfo {pages} {423}
  (\bibinfo {year} {1957})}\BibitemShut {NoStop}%
\bibitem [{\citenamefont {Inogamov}(1992)}]{inogamov}%
  \BibitemOpen
  \bibfield  {author} {\bibinfo {author} {\bibfnamefont {N.~A.}\ \bibnamefont
  {Inogamov}},\ }\href@noop {} {\bibfield  {journal} {\bibinfo  {journal} {JETP
  Lett}\ }\textbf {\bibinfo {volume} {55}},\ \bibinfo {pages} {521} (\bibinfo
  {year} {1992})}\BibitemShut {NoStop}%
\bibitem [{\citenamefont {Abarzhi}\ \emph {et~al.}(2003)\citenamefont
  {Abarzhi}, \citenamefont {Nisihara},\ and\ \citenamefont {Glimm}}]{ang}%
  \BibitemOpen
  \bibfield  {author} {\bibinfo {author} {\bibfnamefont {S.~I.}\ \bibnamefont
  {Abarzhi}}, \bibinfo {author} {\bibfnamefont {K.}~\bibnamefont {Nisihara}}, \
  and\ \bibinfo {author} {\bibfnamefont {J.}~\bibnamefont {Glimm}},\
  }\href@noop {} {\bibfield  {journal} {\bibinfo  {journal} {Phys Lett A}\
  }\textbf {\bibinfo {volume} {317}},\ \bibinfo {pages} {470} (\bibinfo {year}
  {2003})}\BibitemShut {NoStop}%
\bibitem [{\citenamefont {Hill}\ \emph
  {et~al.}(2019{\natexlab{b}})\citenamefont {Hill}, \citenamefont {Bhowmich},\
  and\ \citenamefont {Abarzhi}}]{hillba}%
  \BibitemOpen
  \bibfield  {author} {\bibinfo {author} {\bibfnamefont {D.~L.}\ \bibnamefont
  {Hill}}, \bibinfo {author} {\bibfnamefont {A.~K.}\ \bibnamefont {Bhowmich}},
  \ and\ \bibinfo {author} {\bibfnamefont {S.~I.}\ \bibnamefont {Abarzhi}},\
  }\href@noop {} {\bibfield  {journal} {\bibinfo  {journal} {arXiv}\ ,\
  \bibinfo {pages} {1903.08151}} (\bibinfo {year}
  {2019}{\natexlab{b}})}\BibitemShut {NoStop}%
\end{thebibliography}%

\end{document}